\documentclass[amsfonts,amssymb,amsmath,prl,twocolumn]{revtex4}
\usepackage{times}
\usepackage{graphicx}
\usepackage{bm}
\usepackage{color}

\begin{document}

\title{Degenerate ground state and quantum tunneling in rotating condensates }

\author{Qiang Du$^1$, Martine Le Berre$^2$ and Yves Pomeau$^3$}

\affiliation{$^1$Department of Mathematics, Pennsylvania  State University, PA 16802, USA, \\ $^2$ Laboratoire de Photophysique Mol\'eculaire, Bat.210,
91405 Orsay, France \\ $^3$ Laboratoire de Physique Statistique de l'Ecole normale
sup\'erieure,  24 Rue Lhomond, 75231 Paris Cedex 05,
France.}

\begin{abstract}
Quantum tunneling introduces a fundamental difference between
classical and quantum mechanics. Whenever the classical ground
state is non-unique (degenerate), quantum mechanics restore
uniqueness thanks to tunneling. A condensate in
 a rotating trap with a vortex can have such a degenerate {\emph{classical}} ground state, a degeneracy
 that is excluded in the absence of rotation at least when the Gross-Pitaevskii equation applies. If the rotating
 trap has a center of symmetry, like a figure eight (a peanut), the vortex may be on either side with the same energy
 yielding a degenerate  ground state, a degeneracy lifted by quantum tunneling.  We explain how to compute the rate
 of tunneling in the WKB limit by estimating the action of the trajectory in the Euclidean version of the dynamics.

\end{abstract}

\maketitle

\date{\today }

\section{Introduction}
The search for physical "mesoscopic" situations where quantum tunneling could be significant is motivated by the search
of quantum systems where off-diagonal correlations typical of  quantum mechanics could be created and yield interferences
 without classical counterpart. We introduce below a novel physical system where
such a tunneling exists. As well-known \cite{aft} Bose-Einstein
condensates (BEC) rotating  fast enough have vortices in their
ground state. We find that such vortices may have two different
{\emph{classical}} equilibrium positions mapped into each other by
geometrical symmetry. Later we introduce the basic equations for
describing quantum tunneling between the two degenerate states.
This system is interesting because the Euclidean action allowing
to estimate
 the tunneling frequency is complex, not purely imaginary contrary to cases discussed in textbooks, and the single
  vortex
 of the "classical" state becomes, under the barrier, a pair of interacting vortices, something unnoticed before to the
 best of our knowledge.

The wave-function common to all atoms in a BEC trapped in a potential $V(\bf{r})$ is a solution of the Gross-Pitaveskii
 (GP) equation. This is a function in the ordinary sense because the quantum fluctuations can be neglected in the dilute
 limit, as shown long ago by Bogoliubov. Including in the G-P
equation the possibility of constant rotation of vector ${\bf{\Omega}}$ one has:
  \begin{equation}
i \hbar \frac{\partial \Psi }{\partial t} =
- \frac{\hbar^2}{2m} \Delta \Psi - {i\hbar}
 {\bf{r}} \times {\bf{\Omega}} \cdot\nabla \Psi
 + g|\Psi|^2 \Psi + \widetilde{V}({\bf{r}}) \Psi
 \mathrm{.}
 \label{eq:GP}
\end{equation}
The trapping potential $\widetilde{V}({\bf{r}})$ grows
sufficiently fast at infinity, $g$ is the coupling constant
assumed positive and proportional to the s-wave scattering length,
and $m$ is the mass of each particle. A numerical version of
equation (\ref{eq:GP}) is found by taking as length scale
$\frac{\hbar}{\epsilon \sqrt{mgn}}$, with $n$ being the mean
density of particles over the surface $\sigma$ of the 2D
condensate, $n=\frac{1}{\sigma}\int{|\Psi({\bf{r}})|^2d{\bf{r}}}$,
and $\epsilon$ a dimensionless parameter, arbitrary for the
moment. The unit for ${\bf{\Omega}}$ is $\frac{g n
\epsilon^2}{\hbar}$, and the time unit is $\frac{\hbar}{\epsilon^2
gn}$.  With this system of units, equation (\ref{eq:GP}) becomes
\begin{equation}
i\frac{\partial \Psi }{\partial t} =  -\frac{1}{2} \Delta \Psi + i
 {\mathcal{M}}  \Psi + \epsilon^{-2} \left(|\Psi|^2 \Psi + V({\bf{r}})
\Psi\right) \mathrm{,} \label{eq:GPscaled}
\end{equation}
where  ${\mathcal{M}}$ is the real
antisymmetric linear operator $ {\mathcal{M}}  = \omega \left(x
\frac {\partial}{\partial y}-  y \frac {\partial}{\partial
x}\right) $, with $(x,y)$ being rectangular coordinates normal to the
rotation axis and $\omega=|{\bf{\Omega}}|\frac{\hbar}{gn\epsilon^2}$.
Furthermore the potential $V({\bf{r}})$ has  been made
dimensionless.
 The ground state is a solution of (\ref{eq:GP}) in the form
of $e^{-i\mu t}\Psi(\bf{r})$ with  the smallest $\mu$, for the
given $ \int {\mathrm{d}} {\bf{r}}  |\Psi|^2$. Without rotation (${\bf{\Omega}} = 0$),  the ground state is non-degenerate
\cite{aft}, but no such general result exists with rotation, and we give below  examples of the opposite.

Consider a potential $V({\bf{r}})$ such that the condensate is constrained to stay inside a closed curve shaped like
 a peanut.
The idea behind this choice is that the vortex created by rotation
may have an equilibrium position on either side of the peanut and
so yield a degenerate ground state. We get such a curve from
the Cassini oval, the locus of a point at which the product,
$b^2$,
 of the distances to two foci of coordinates ${\bf r} = (\pm a,0)$,
 is constant. The oval is made of two pieces for $b<a$ and of a single one for
$b>a$ \cite{J.D. Lawrence}.
 Physically the length $a$ (for instance) is such that the dimensionless 
parameter
 $\epsilon$ is small. The Cartesian equation of the oval is
$V({\bf{r}}) = 0$ with
$$V({\bf{r}}) =  b^4 + 4a^2 x^2  -  (x^2 + y^2  + a^2)^2\mathrm{.} $$
 Besides $\epsilon$, two dimensionless
parameters characterize the system, one geometrical, the ratio
$\frac{b}{a}$ (assumed to be bigger than $1$), and the other measuring the strength of the perturbation due to rotation.
Physically, this strength is found by comparing the circulation around a vortex and the one due to a solid body
rotation at angular velocity $\omega$. This yields the dimensionless ratio
  $\omega = \frac{m |{\bf{\Omega}}| a^2}{\hbar}$, and defines $\epsilon=\frac{\hbar}{a\sqrt{ngm}}$. One expects a
   transition for a finite value of
  $\omega$ from a vortex-free ground-state to a ground-state with vortex, a transition depending on the value of $\frac{b}{a}$.
To find the ground state,
we may simply consider the minimizers of the following nondimensionalized GP-functional, or energy:
 \begin{equation}
\int_D \{|\nabla \Psi |^2 - 2\Re(i\Psi {\mathcal{M}} \bar{\Psi})
+
\frac{1}{2\epsilon^2}|\Psi|^4 + \frac{V({\bf r})}{\epsilon^2}|\Psi|^2
\} d\bf{r}
 \mathrm{,}
 \label{eq:GPf}
\end{equation}
where
 $D$ is
the domain defined by $V({\bf r})>0$ and $\Re(\cdot)$ denotes
the real part of its argument, $\bar{\Psi}$ being the complex conjugate of $\Psi$.
This minimum is constrained by
 the normalization condition
$$
\int_D |\Psi|^2 d{\bf{r}}= \int_D V^+({\bf r}) d{\bf{r}}\mathrm{,}
$$
 where $ V^+({\bf r})$ denotes the positive part of the
Cassini oval potential $V({\bf r})$. Different solution branches
are found for a range of the parameter values of $a,b,\epsilon$
and the rotation speed $\omega$ using numerical methods similar to
that described in \cite{Adu}. We hereby present some examples.

\begin{figure}[htbp]
\centerline{$\;\;$
\includegraphics[height=1.7in]{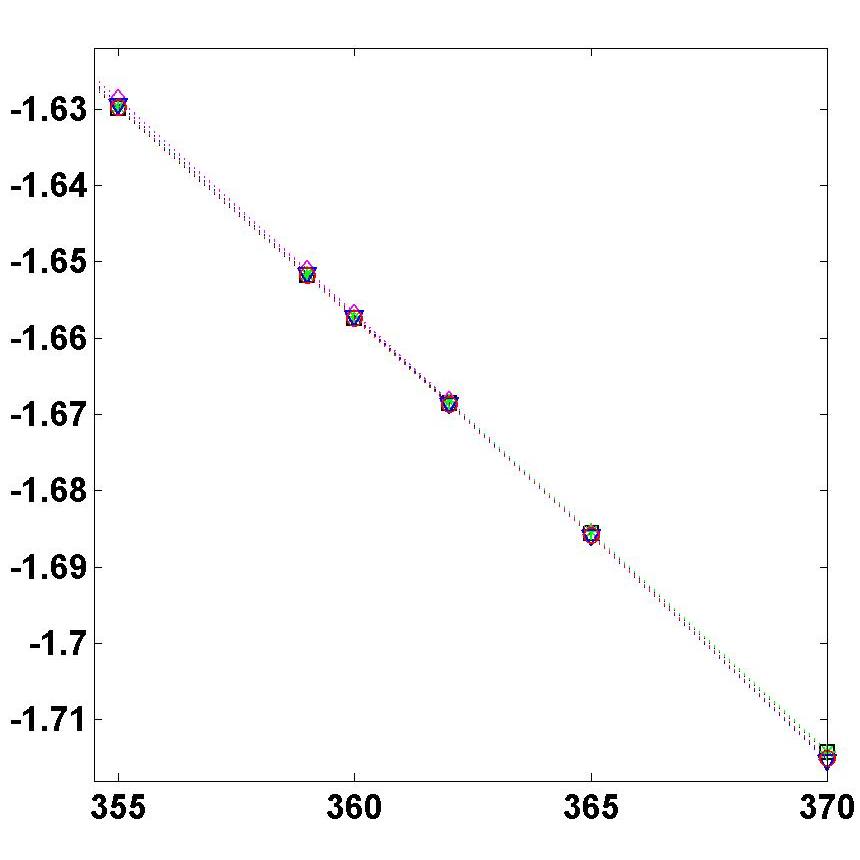}
\includegraphics[height=1.7in]{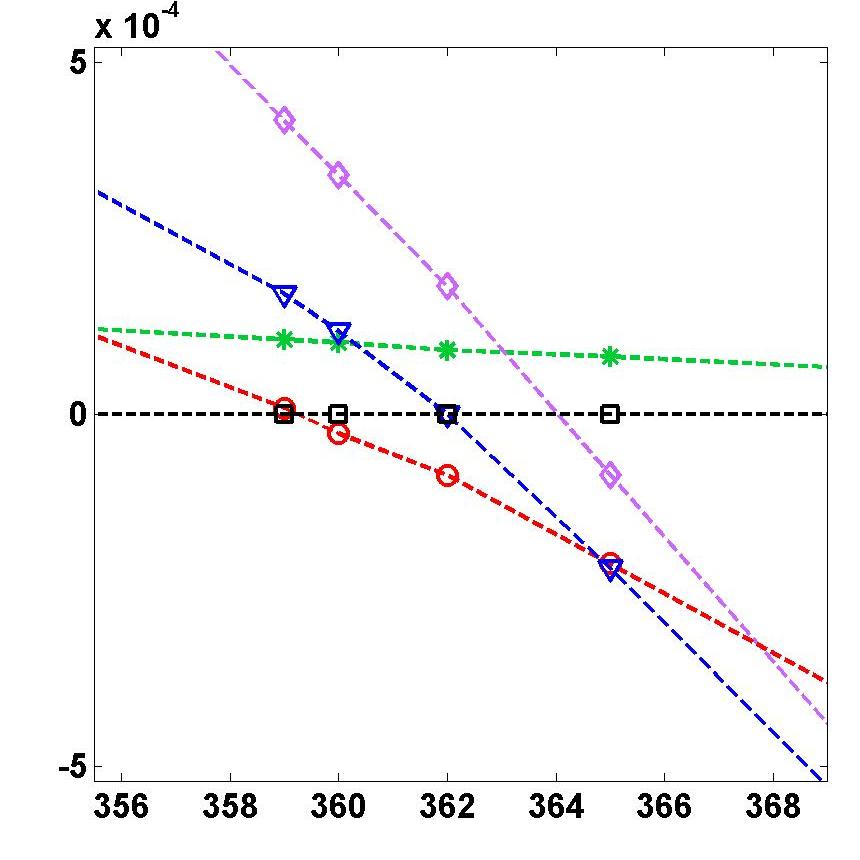}$\;\;$}
\caption{Energy values  (left) and relative
energy differences (right),
as functions of the rotation speed, for different solutions
with the Cassini oval potential.
Black-square: no vortex. Red-circle: non-symmetric vortex.
Blue-triangle: two vortices. Purple-diamond: three vortices,
Green-star: one symmetric vortex. } \label{e275}
\end{figure}

\begin{figure}[htbp]
\centerline{
\includegraphics[height=0.7in]{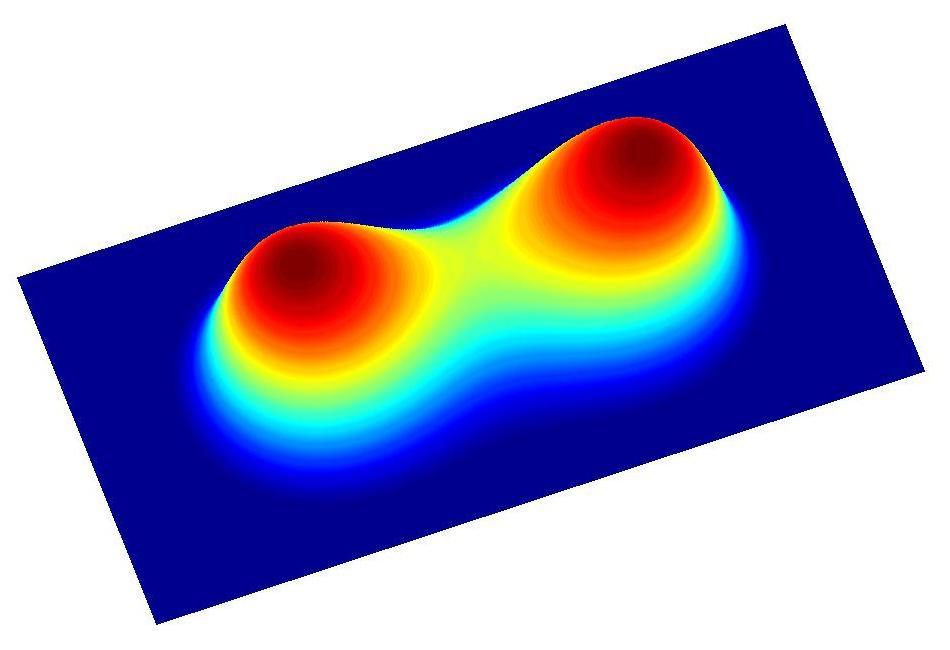}
\includegraphics[height=0.7in]{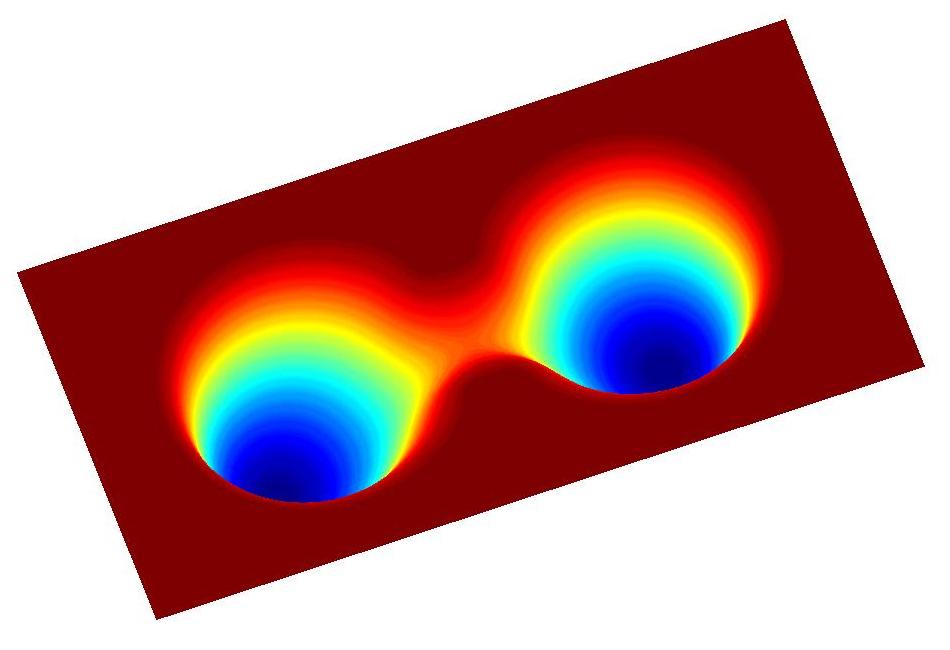}
\includegraphics[height=0.7in]{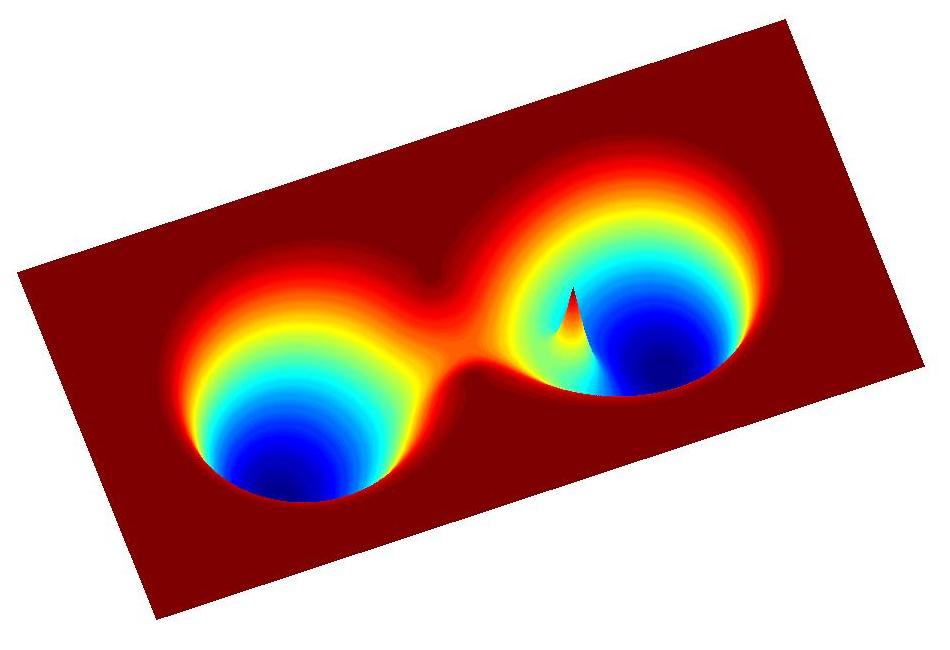}}
\centerline{
\includegraphics[height=0.7in]{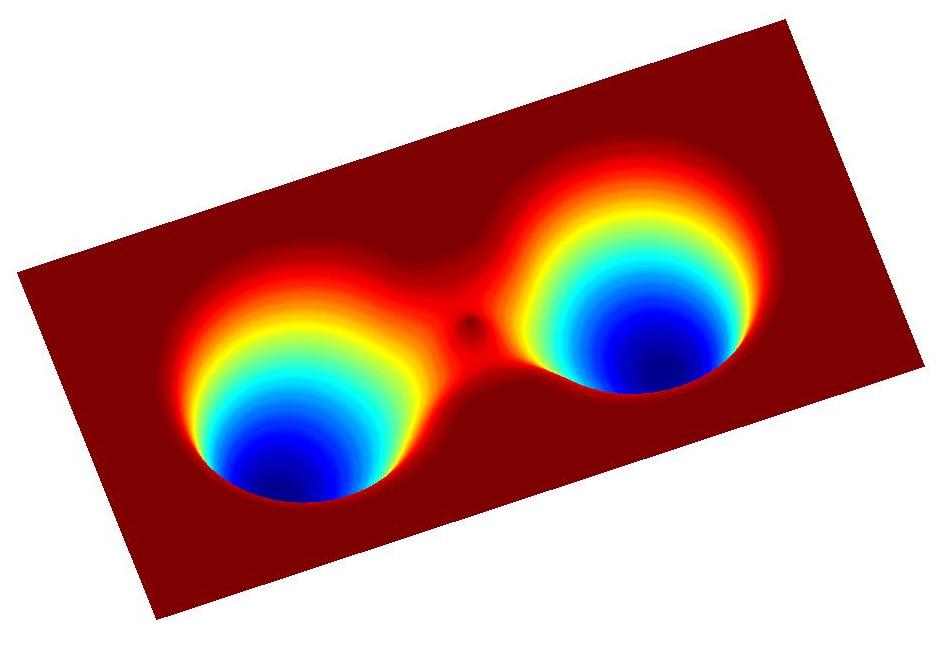}
\includegraphics[height=0.7in]{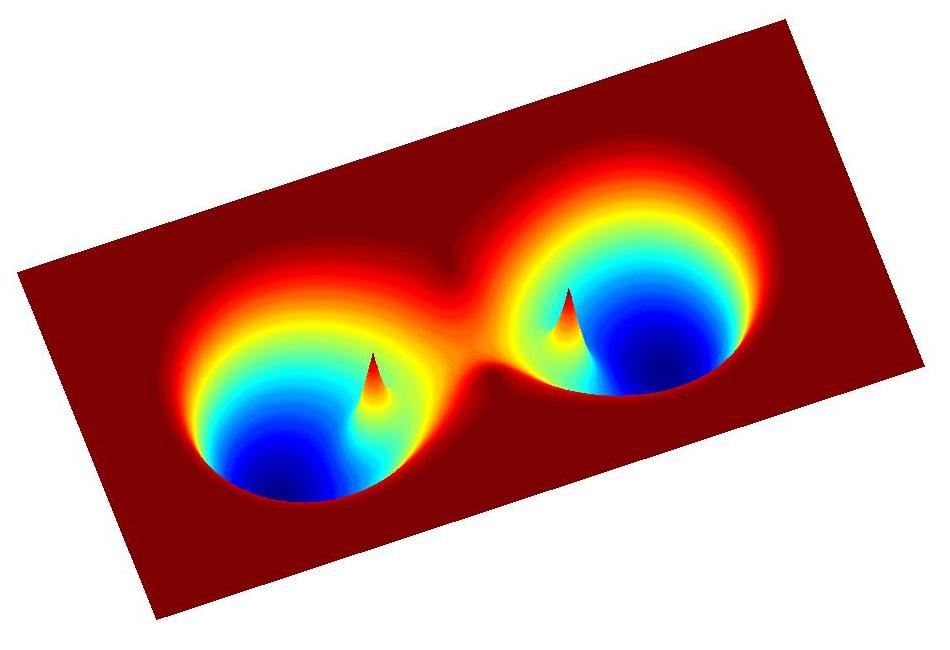}
\includegraphics[height=0.7in]{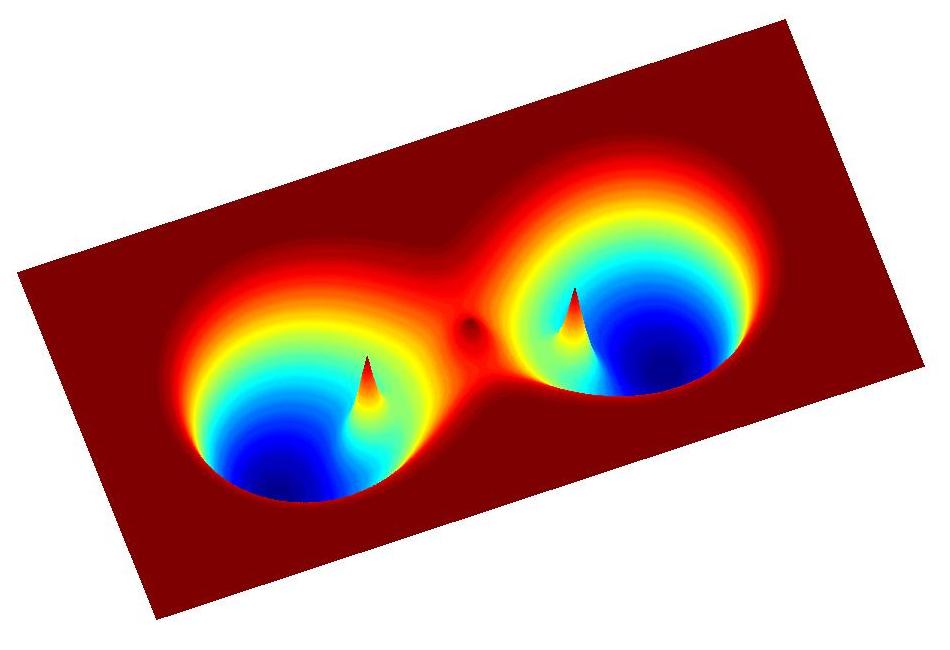}}
\caption{Surface plots of $|\Psi|$ corresponding to the ground
state without rotation (first row left, top view) and the five
solutions found at the rotation speed $\omega=360$ (upside-down
view, they are respectively solutions without any vortices in the
interior oval, with a single non-symmetric vortex in one half of
the oval, with a vortex in the center of the oval, with two
symmetric vortices, and with three vortices). At this rotation
speed, the ground state is the one with a non-symmetric vortex
which is a degenerate state together with its mirror image with
respect to the y-axis. } \label{fh360f275}
\end{figure}

For $a=0.25,b=0.275$, and $\epsilon=0.001$, Fig.\ref{e275}
displays the energy  computed using
(\ref{eq:GPf}) for the various solution branches (i.e. the
solutions without vortex, with one symmetric vortex, with one
off-center  vortex, with two symmetric vortices, and with three
vortices) as a
function of the rotation speed. The relative energy differences
$(1-\frac{E_{n}}{E_{0}})$ are highlighted for $\omega$ near $360$,
where $E_{n}$ is the energy of a solution with at least one
vortex,  and $E_{0}$ the energy of the vortex free solution
corresponding to the same rotation speed. Although the differences
are very small, it is clear that in this range of $\omega$, the
solution
 with a non-symmetric (off-center) vortex has the lowest energy and makes the ground state.
Therefore our initial guess is correct: in this range of parameters values, the ground state with an off-center
vortex is degenerate.
The different solutions are shown in Fig.\ref{fh360f275}.

\begin{figure}[htbp]
\centerline{$\;\;$
\includegraphics[height=1.7in]{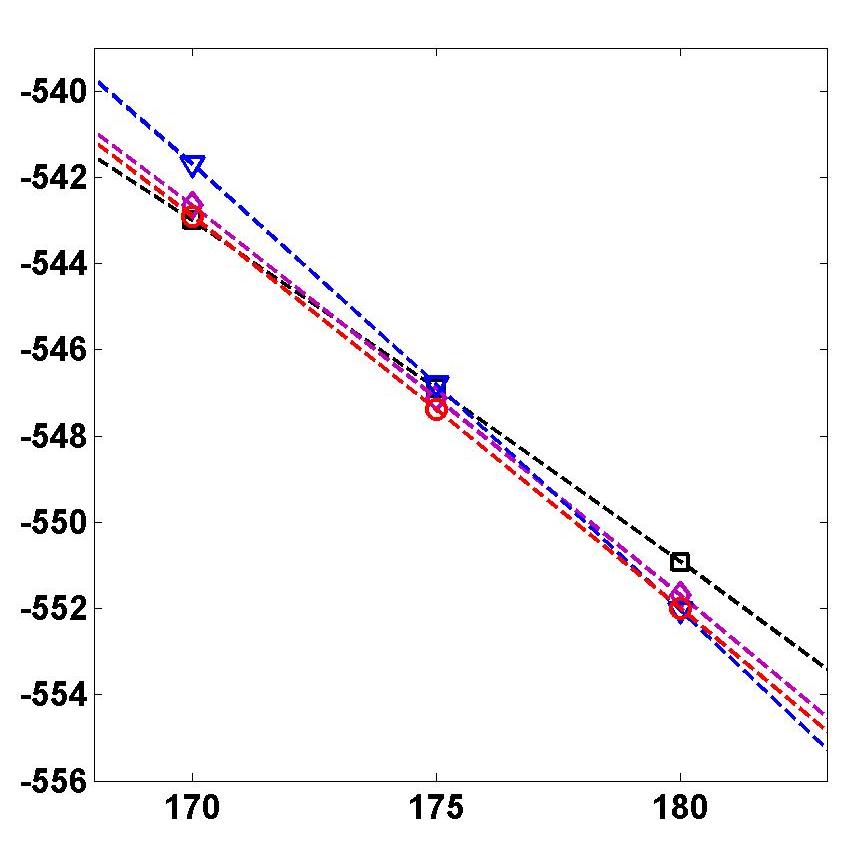}
\includegraphics[height=1.7in]{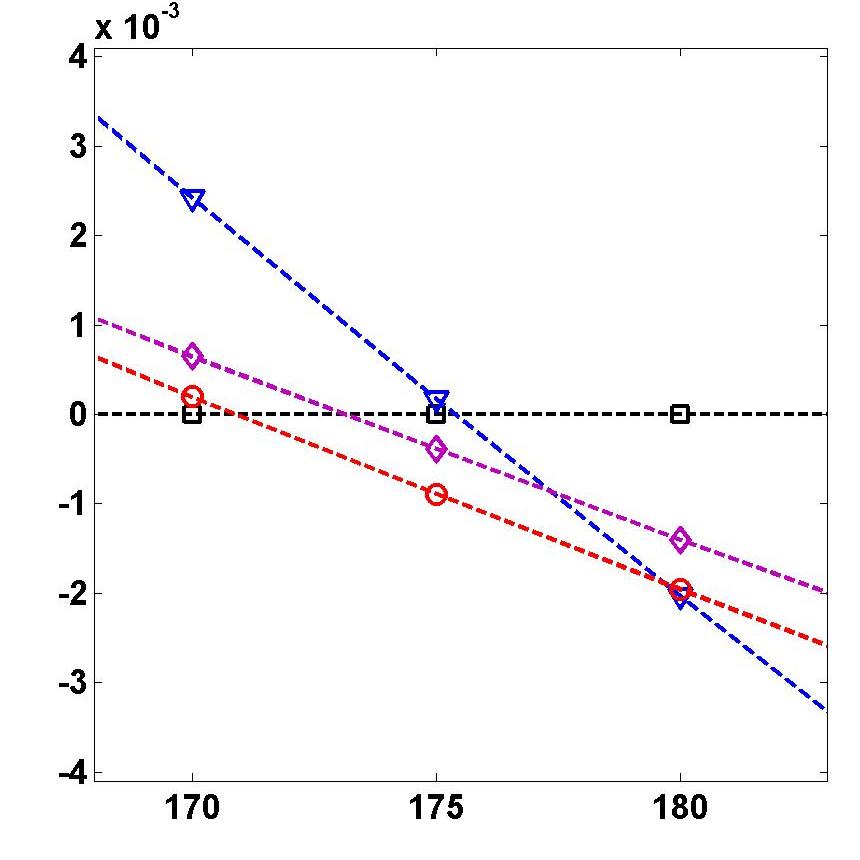}
$\;\;$}
\caption{Energy values  (left) and
relative energy differences (right),
as functions of the rotation speed,
for different solutions with a piecewise constant
potential.
Black-square: no vortex. Red-circle: non-symmetric vortex.
Blue-triangle: two vortices. Purple-diamond:
 one symmetric vortex.}
\label{sqae}
\end{figure}

\begin{figure}[htbp]
\centerline{
\includegraphics[height=1.0in]{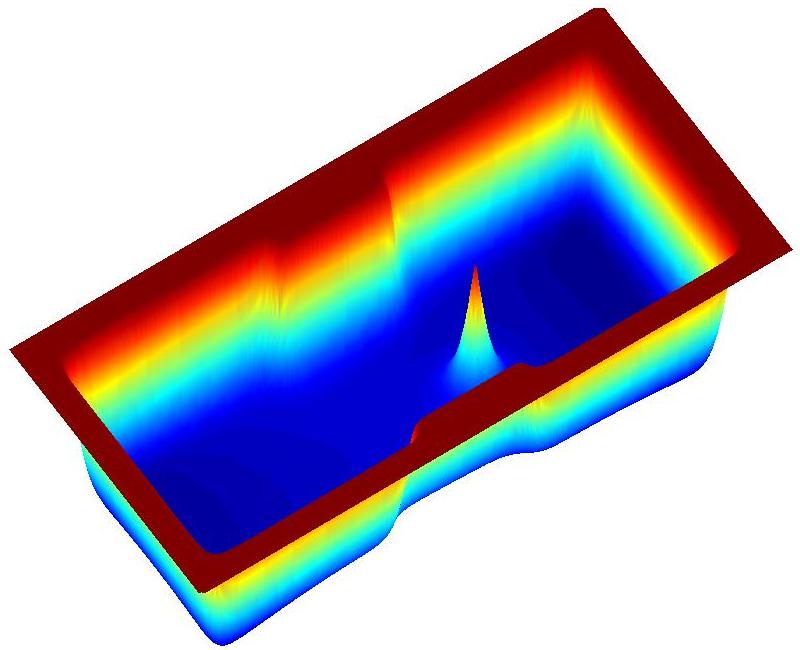}}
\caption{ Plot of $-|\Psi(x,y)|$ for a non-symmetric degenerate
ground state with the piecewise constant potential at the rotation
speed $\omega=175$.} \label{sqa}
\end{figure}

Similar energy diagrams can be found for different values of the
parameters. With a smaller $b=0.27$ but the same values of $a$ and
$\epsilon$, only three different solutions branches were found
near $\omega = 415$.
We expect that a degenerate ground-state with
a non-symmetric vortex exists in other geometries than the Cassini
oval. We have replaced the Cassini oval potential $V({\bf r})$
 by a piecewise constant potential of value $+1$
inside two squares connected through a rectangular channel,  and
$(-10)$ elsewhere. When the channel width is that of the squares,
$D$ is a rectangle, and the ground state solutions keep the domain
symmetry. A non-symmetric
vortex branch appears for a smaller channel width. We found that
for a narrower channel the energy differences of the different
solution branches are larger than for the Cassini oval when the
ground state is degenerate (see Figs.\ref{sqae} and \ref{sqa}).
Even larger energy differences between the non-symmetric ground
state and the other solutions are found by decreasing the potential 
in the channel connecting the two sides.

Based on our simulations, the following observations are in order:
by increasing rotation speed with a standard symmetric harmonic
potential, the ground state is given first \cite{aft} by the vortex
free solution, then the single centrally symmetric vortex
solution, then the solution with two and/or more vortices. It can be
argued that, if in the range of the rotation speed where the
ground state has only one vortex, there are substantial energy
differences between this ground state and other solutions (such as
the vortex-free and the multi-vortex solutions), by properly
perturbing such potentials in a non-symmetric fashion,
non-symmetric ground state can be created and some can be
degenerate. For instance, by lowering the potential away from the
center (but not too far), symmetrically with respect to the
$y$-axis, a couple of local {\em pinning sites} can be formed which
would attract the central vortex to the vicinity of one of these
sites, and lead to a solution whose energy is lower than the
central vortex state. Yet if the perturbation is small, then the
changes to the energies of the vortex free or multi-vortex
solutions are also small, hence leaving the non-symmetric vortex
as the new ground state. Such a solution can have the vortex
situated at either one of the two pinning sites, thus becomes
degenerate. This scenerio has been confirmed in our
numerical experiments as well. Other configurations of pinning sites, with
three or four-fold symmetry for instance, could produce solutions
with  even higher
degeneracy. The keys to realize this experimentally, aside for being able
to produce the right geometric pattern, are to increase enough the
pinning strength to differentiate the energy between the central vortex state
and the pinned vortex state and to, at the same time,  maintain substantial energy differences between the single vortex solution and the other solution
branches. Our simulations provide strong evidence to such
possibilities. It did not escape our attention that a superconducting thin film in a perpendicular magnetic field
 yields a closely related situation, mathematically speaking at least \cite{dingdu1}. Therefore we expect that in certain range of parameters too the same kind of peanut like geometry yields there a degenerate ground state with a non-symmetric vortex.

Whenever the classical ground-state is degenerate, one expects quantum tunneling to restore the symmetry and make
 one single {\emph{quantum}} ground-state out of the set of different {\emph{classical}} ground-states derived from
 each other by symmetry. In the semi-classical approximation, the tunneling time $T_Q$ is proportional
 to $\exp(S_{E}/\hbar)$, where $S_{E}$ is the Euclidean
 classical
 action, supposed to be much larger than $\hbar$.
 If the system is very large, $T_Q$ may be so large that quantum tunneling is never observed. Nevertheless one may imagine
  intermediate cases, outside of the realm of atomic or
  nuclear phenomena, where the Euclidean action is not
  that large compared to $\hbar$ , leading to a non-negligible tunneling probability. The tunneling of an optical soliton
  between two coupled fibers \cite{MartYv} yields such an example.

Let us discuss the relevance of the various possible {\em tunneling} effects \cite{PomSim}. We mean now by {\em tunneling}
an effect allowing a physical system in a double well potential to go from one well to the other either by purely
quantum effects or by a mixture of thermal and quantum fluctuations or by thermal fluctuations only. This defines
 an effective tunneling time. The first typical time is the quantum tunneling time, $T_Q$, then the coherence time
  $T_c$. The latter time is the maximum time during which the quantum state inside one of the wells remains coherent,
  supposing that the condensate is at finite temperature, since at zero temperature this coherence time is infinite.
   In the case of BEC, $T_c$ is the mean-free flight time of a particle in the ground state colliding with a
   thermal particle divided by the number of particles in the ground state, the collision process being Poissonian.
    If there are many particles in the ground state, this coherence time may become very short because of this division.
     The third typical time is the Arrhenius-time $T_A$ for the passage over the energy barrier separating the two classical ground states under the effect of the thermal fluctuations. If $T_A$ is very large, two possibilities are met. If $T_Q \ll T_c$, the effective tunneling time is $T_Q$. If $T_c \ll T_Q$, the effective tunneling time is of order $\frac{T_{Q}^2}{T_c}$  \cite{PomSim}. If the Arrhenius time is less than $T_Q$, the transition will always be by activated barrier crossing. Otherwise, the shortest of the two times,  $\frac{T_{Q}^2}{T_c}$ or $T_A$, will be the effective transition time.

In the present problem the tunneling time can be calculated by
quantizing the dynamical GP equation \cite{noteuv}. If one does
not attempt to compute the prefactor of the exponential
$\exp(S_{E}/\hbar)$, the derivation of $T_Q$, that amounts to
calculate
 the Euclidean trajectory, is usually fairly standard. However, because the present problem has a peculiarity
   compared to the ones usually treated, we shall explain the general idea in some details.
The equation (\ref{eq:GPscaled}) is the Euler-Lagrange condition making stationary the action
$$
{\mathcal{S}} = \int {\mathrm{d}}{\bf{r}}  {\mathrm{d}}{t} \mathcal{L}
 \mathrm{,}
$$
associated to the Lagrangian $\mathcal{L}$, which is the following
functional of $\Psi$ and $\overline\Psi$,
\begin{eqnarray}
&& {\mathcal{L}} = \frac{i}{2}\left(\Psi \frac {\partial \overline{\Psi} }{\partial t}  - \overline{\Psi} \frac {\partial{\Psi} }{\partial t}\right) + \frac{1}{2}|\nabla\Psi|^2 + \frac{1}{2\epsilon^2} |\Psi|^4 \nonumber\\
&&\qquad +\frac{1}{\epsilon^2} (V({\bf{r}}) - \mu)  |\Psi|^2  +
\frac{i}{2} \left[\overline{\Psi} {\mathcal{M}}\Psi - {\Psi}  {\mathcal{M}}
\overline{\Psi}\right]
 \mathrm{.}\;
 \label{eq:GPELpsi}
\end{eqnarray}

We include the ground state energy $\mu$ in (\ref{eq:GPELpsi})
to make it clearer the possibility of real or complex
  solutions.
To perform the analytic continuation of the dynamical equations,
one has to write first the relevant quantities with real
functions. Let us write $\Psi$ as $\Psi = u + iv$ with $u$ and $v$
being real
functions of $\bf{r}$ and $t$. The Lagrangian (\ref{eq:GPELpsi})
becomes:
$$
{\mathcal{L}} = (u \frac {\partial v }{\partial t}  - v \frac
{\partial u }{\partial t} ) + {\mathcal{L_{1}}}\mathrm{,}
$$
where

\begin{eqnarray}
&&{\mathcal{L_{1}}}= \frac{1}{2} \left( (\nabla u )^2 + (\nabla v
)^2 \right)+ \left[ v \mathcal{M} (u) - u \mathcal{M} (v) \right]
 \nonumber \\
&&\qquad +\frac{1}{\epsilon^2}(V({\bf{r}})-\mu )(u^2 + v^2)+
\frac{1}{2\epsilon^2} (u^2 + v^2)^2\mathrm{.}\;  \label{eq:GPELuv2}
\end{eqnarray}

In Feynman's formulation of quantum mechanics \cite{Feynman} any
observable is given by a formal integral over all paths with an
integrand that is a function times an exponential $e^{i
\frac{{\mathcal{S}} (X(\cdot))}{\hbar}}$, ${\mathcal{S}} (X(\cdot))$ being
the classical action of the path $X(\cdot)$. Here
${\mathcal{S}(X(\cdot))}$ would be calculated along all paths where
the two functions $u(t,\bf{r})$ and $v(t,\bf{r})$ represent the
vortex on one side (at time minus infinity) and the other with the
vortex on the other side (at time plus
   infinity). Along the path, the wave-function $\Psi({\bf{r}},t))=u({\bf{r}},t)+iv({\bf{r}},t)$ is not
necessarily a
 solution of the dynamical equation (\ref{eq:GP}), but must tend at
 $ t=\pm\infty$ to steady
 solutions of the GP equation (note that $\mu \Psi$ is subtracted in equation (\ref{eq:GP})
 to have a time independent solution). The tunneling corresponds to situations where the end-points of the trajectory
  $X(\cdot)$ are not linked by a
    classical trajectory, yielding the dominant saddle-point contribution to Feynman's integral at $\hbar$ small.
In this limit, and if there is no such classical trajectory, one
extends the range of possible
 values of $t$ to the complex plane. The Euclidean action associated to tunneling is found by taking $t$ purely imaginary
 in the equations of motion,  $t =
i\tau$, $\tau$ real, leading to the expression
\begin{equation}
 {\mathcal{L}}_{E} = i\left(v \frac {\partial u }{\partial \tau }
- u \frac {\partial v}{\partial \tau }\right) + {\mathcal{L_{1}}}
 \mathrm{.}
 \label{eq:euclLag}
\end{equation}

The Euclidean equations of motion for $u$ and $v$ are derived by
variation of $\int dt \int d{\mathbf{r}}{\mathcal{L}}_{E}$,
leading to  the two coupled equations:
\begin{equation}
i \frac {\partial u}{\partial \tau} = -\frac{1}{2}\nabla^2 v  +
 \mathcal{M}(u) + \frac{v}{\epsilon^2}(u^2 + v^2+V - \mu)
 \mathrm{,}
 \label{eq:GPuEu}
 \end{equation}
and
 \begin{equation}
i \frac {\partial v}{\partial \tau} = \frac{1}{2}\nabla^2 u -
\mathcal{M}(v) -\frac{u}{\epsilon^2}(u^2 + v^2+V - \mu)
 \mathrm{.}
 \label{eq:GPvEu}
 \end{equation}

 If the
 contribution proportional to $\mathcal{M}$ is absent
(that is without rotation), the Lagrangian in equation  (\ref{eq:GPELuv2}),
 and the dynamical equations (\ref{eq:GPuEu})-(\ref{eq:GPvEu}) become real
 by changing $v$
 into $iv$, yielding a formally real Euclidean dynamics.
  However this cannot be done with a non-zero ${\mathcal{M}}$ because the
 transformed Lagrangian is complex and cannot be made real by any simple transformation. Although this is rarely
 considered, a fully complex Euclidean action does not hurt any general
 principle.

 In the present problem, the tunneling probability is proportional to $\exp(\Re({S_{E}})/\hbar)$, and the end
 states,
 at times
$\tau=\pm  \infty$, are the classical ground states, with real
$u_\infty$ and $v_\infty$, because
 either in real or in Euclidean dynamics, the equilibrium
state is the state which minimizes $\int
d{\mathbf{r}}{\mathcal{L_1}}$ (i.e. the part of the action
 not proportional to time derivatives).
 Note that by substituting $t$ for $-i\tau$  in the pair of equations (\ref{eq:GPuEu}) and  (\ref{eq:GPvEu}),
 one recovers the classical  GP equation, where $u$ and $v$ are purely real. In the case considered here, where
 the asymmetrical one-vortex state is the ground state, the GP equation leads to stable equilibrium states,
  ($u_\infty$ and $v_\infty$), that forbids any tunneling. While the tunneling occurs
  when using the Euclidean equations
 (\ref{eq:GPuEu})-(\ref{eq:GPvEu}) with real $\tau$, where ($u_\infty$ and $v_\infty$) are unstable equilibrium states.
 For arbitrary $\tau$, the functions $u$ and $v$ are both complex, without any simple
relation in between, like being a pair of
complex conjugates. Therefore
  the present Euclidean problem  involves two complex fields, each having
 their own set of {\emph{different}} vortices, a rather
unusual fact.
 The numerical integration of the Eulerian system
requires some caution, and will be treated in a future
publication. Actually, while the mathematical problem is well
posed, as a \textit{Dirichlet problem} with boundary conditions at
infinite time, the numerical search of solutions should avoid to
solve an initial value \textit{Cauchy problem} that is ill-defined because
of the instability of this Euclidean dynamics at very short
wavelength \cite{largek}. Furthermore, a change of
variable $\tau$ into, for example $\tanh(\tau)$, would lead to a
finite integration time. Then the trajectory joining the two
equilibrium states, could be found by starting slightly off one
equilibrium state and by finding the pair 
 $u,v$ that minimizes  $|\int
d{\mathbf{r}} d{\tau} {\mathcal{L}}(u,v)|^2$. This
procedure should avoid the instability effects for
large wave numbers. Along this
trajectory, the transient vortices formed on each field, will
merge {\em at the end}, into a single one.

\begin{acknowledgments}
The work was initiated while the authors Du and Pomeau were visiting the Institute for
Mathematical Sciences, National University of Singapore in 2007. The visit was supported
by the Institute.
The research of Du is also supported in part by the NSF grant DMS-0712744.
\end{acknowledgments}

{}


\thebibliography{9}

\bibitem{aft} A. Aftalion, \emph{Vortices in Bose-Einstein condensates}, {Birkhauser, Boston (2006)}.

\bibitem{J.D. Lawrence} section 5.16 in J.D. Lawrence, {\emph{A catalog of special
plane curves}}, Dover publication, New-York (1972).

\bibitem{Adu} A. Aftalion and Q. Du,
Phys. Rev. A, {\textbf{64}}, 063603 (2001).
A. Svidzinsky and A. Fetter, Phys. Rev. Lett.  {\textbf{84}}, 5919 (2000).


\bibitem{dingdu1}
S. Ding and Q. Du, SIAM J. Math. Anal., {\textbf{34}}, 239 (2002).
S. Ding and Q. Du,
Comm. Pure Appl. Anal., {\textbf{1}}, 327 (2002).

\bibitem{Feynman} R. P. Feynman, R. B. Leighton and M. Sands,
{\emph{Lectures on Physics}}, Add-Wesley Publ., Readings (1964).

\bibitem{PomSim}Y. Pomeau and  A. Pumir,
J. de Phys. (Paris) {\textbf{46}}, 1797 (1985).

\bibitem{noteuv} There exists a formally straightforward way of quantizing the
vortex dynamics, by taking the cartesian components of the vortex
position, ($x,y$), as conjugate variables (the so-called Kelvin's dynamics). Generally this is not
equivalent to the quantization of the original GP equation, simply
because the physical action as calculated from the GP
equation, has the same dimension as $\hbar$, that is not the case
when quantizing the Kelvin dynamics.


\bibitem{MartYv} Y. Pomeau and M. Le Berre,
Chaos, {\textbf{17}}, 037118 (2007).

\bibitem{largek} The short-wave (large $k$) approximation yields solution of the coupled set
(\ref{eq:GPuEu})-(\ref{eq:GPvEu}), proportional to $e^{i
{\bf{k}}\cdot{\bf{r}} \pm \epsilon^2 k^2 \tau}$ (when considering
the first term in the r.h.s.). Because of the very fast growth in
"time" $\tau$ of the amplitude of the exponential, the set
(\ref{eq:GPuEu})-(\ref{eq:GPvEu}) makes an ill-posed
{\emph{initial value}} problem. However the unlimited growth in
time of the Fourier mode becomes irrelevant when one considers, as
here, this set as defining an elliptic problem in time, with
boundary condition at plus and minus infinity.
\endthebibliography
\end{document}